\documentclass[a4paper,fleqn,usenatbib]{mnras}
\usepackage{amssymb,amsmath,graphicx,psfrag,mathtools}
\usepackage[T1]{fontenc} %MNRAS
\usepackage{ae,aecompl} %MNRAS
\usepackage{url}
\usepackage{revsymb}
\hypersetup{draft} %TO AVOID ERROR: \PDFENDLINK ENDED UP IN DIFFERENT
\title[Periodically Modulated FRB Activity]{Testing Models of Periodically Modulated FRB Activity}
\author[J. I. Katz]{
J. I. Katz,$^{1}$\thanks{E-mail katz@wuphys.wustl.edu} %MNRAS
\\
$^{1}$Department of Physics and McDonnell Center for the Space Sciences,
Washington University, St. Louis, Mo. 63130 USA %MNRAS
}
\date{Accepted XXX.  Received YYY; in original form ZZZ} %MNRAS
\pubyear{2020} %MNRAS
\date{\today}
\begin{document} %MNRAS
%\psfrag{theta}{$\theta$}  Works only on COMPLETE strings
\label{firstpage} %MNRAS
\pagerange{\pageref{firstpage}--\pageref{lastpage}} %MNRAS
\maketitle %MNRAS
\begin{abstract}
The activity of the repeating FRB 20180916B is periodically modulated with a
period of 16.3 days, and FRB 121102 may be similarly modulated with a period
of about 160 days.  In some models of this modulation the period derivative
is insensitive to the uncertain parameters; these models can be tested by
measurement of or bounds on the derivative.  In other models values of the
uncertain parameters can be constrained.  Periodic modulation of aperiodic
bursting activity may result from emission by a narrow beam wandering within
a cone or funnel along the axis of a precessing disc, such as the accretion
discs in X-ray binaries.  The production of FRB 200428 by a neutron star
that is neither accreting nor in a binary then shows universality: coherent
emission occurring in a wide range of circumstances.
\end{abstract}
 %MNRAS
\section{Introduction}
The activity of the repeating FRB 20180916B has been shown to be strongly
modulated, with all detected bursts confined to about 30\% of an $\approx
16.35\,$d cycle \citep{CHIME20,PM20,Pi20,Pl20}.  There is also evidence for
an $\approx 160$ d period in FRB 121102 \citep{R20,C21}.  This behavior was
unexpected.  Most FRB models are based on neutron stars with predicted
rotational periods ranging from milliseconds (FRB as young energetic pulsars
emitting giant pulses) to seconds (FRB related to Soft Gamma Repeaters
(SGR); confirmed for FRB 200428 identified with SGR 1935$+$2154 with a
period of 3.245 s).  These models do not naturally lead to predictions that
the aperiodic activity of repeating FRB would be periodically modulated with
periods of many days.

A number of models for these long periods were soon proposed.  These fell
into three categories:
\begin{enumerate}
	\item Binary orbital periods, with a wind from one star either
		obscuring radio emission from its companion FRB source or
		interacting with that source \citep{IZ20,LBG20,Po20,ZG20}.
	\item Precession of a strongly magnetized neutron star sweeping
		possible emission directions.  In one version of this model,
		the neutron star precesses freely as a result of a
		magnetically distorted non-spherical figure
		\citep{LBB20,ZL20}.  In other versions it precesses under
		the influence of radiative torques \citep{So20}, the
		Newtonian torque of a fallback disc \citep{TWW20}, or
		relativistically (geodetic precession) in a very short
		period binary \citep{YZ20}.
	\item Rotation of a neutron star source with the observed 16.3 day
		period \citep{BWM20}.
\end{enumerate}
Most of the these models predict that the period will change, and some of
their predictions can be cast in forms that are insensitive to the unknown
parameters, permitting tests of the models.  Sec.~\ref{testing} quantifies
these predictions.  Sec.~\ref{disc} develops the suggestion of \citet{K20}
that long period modulation is the result of precession of the axis of an
accretion disc under the torque of a binary companion star.
\section{Testing the Models}
\label{testing}
It is straightforward to suggest a model that can produce the observed
period, and the Introduction cites several.  It is now two years since this
modulation was first observed \citep{CHIME20}.  New data extend over 50
cycles (a duration of $T = 49 P$): \citet{Pl20} found a period $P = 16.33
\pm 0.12\,\text{d} = 1.411 \pm 0.010 \times 10^6\,$s and \citet{PM20} found a
period $P = 16.29\,{+0.15 \atop -0.17},\text{d} = 1.407\,{+0.013 \atop
-0.015} \times 10^6\,$s, {consistent with each other and with 16.35 d}.
$P$ and the angular modulation frequency $\Omega = 2 \pi/P = 4.46 \pm 0.03
\times 10^{-6}\,\text{s}^{-1}$ are accurately established.

Significant constraints can be placed on the rate of change
\begin{equation}
	\label{Omdot}
	|{\dot \Omega}| = {8|\Delta\phi| \over T^2} \lessapprox 5 \times
	10^{-16}\ \text{s}^{-2},
\end{equation}
where it is assumed the period (or $\Omega$) is fitted to data from an
entire data interval $T$; $|\Delta\phi| \lessapprox 0.05\,\text{cycle}
\approx 0.3\,$radian is the phase deviation at the endpoints from exact
periodicity at a period fitted to the midpoint of the data.  As $T$
increases the constraint (\ref{Omdot}) will rapidly become stricter or a
significant non-zero $\dot \Omega$ will be observed.

The following subsections address each proposed origin of the 16.3 d period.
Some may be excluded by Eq.~\ref{Omdot}; others will be tested in the future
as $T$ increases.
\subsection{Orbital Period}
Orbital periods are expected to be extremely stable, and not usefully
constrained by Eq.~\ref{Omdot}.  However, LOFAR observed \citep{Pl20,PM20}
bursts from FRB 20180916B at frequencies as low as 110 MHz over the same or
a larger fraction of the cycle as at the much higher frequencies observed by
CHIME/FRB and Apertif.  These authors pointed out that this argues against
binary models in which a wind from the companion, more opaque at lower
frequencies, limits the orbital phases at which a FRB can be observed.
{\citet{IZ20} and \citet{LBG20} propose different physical origins of
this opacity.  The resulting predicted phenomenologies are similar.}

The phase delay (LOFAR detections occurring later in the 16.3 d cycle) might
be explained by a reflecting or refracting wind whose flow is bent by the
emitting neutron star's gravity.  If this carves out a {transparent} low
density channel, {as depicted by \citet{IZ20,LBG20},} it might direct
the low frequency radiation as did the Holmdel Horn Antenna \citep{CHH61}
with which Penzias and Wilson discovered the 3 K background radiation.

A plasma frequency of 50 MHz, sufficient to reflect (at grazing incidence)
radiation in the LOFAR band, corresponds to an electron density $n_e \sim 4
\times 10^7\,\text{cm}^{-3}$.  For a plausible dimension $\sim 10^{12}\,$cm
of a 16.3 d orbit this would imply a variation of dispersion measure,
periodic with the orbital period, $\Delta\text{DM}\sim 10\,$pc-cm$^{-3}$ of
the higher frequency radiation that would penetrate such a plasma; this is
not observed.  If the plasma were dense enough to reflect the higher
frequencies, there would be no phase delay between these bands.
\subsection{Neutron Star Precession Period}
{A number of models have been proposed in which the observed modulation
period is the precession period of a much faster rotating neutron star.  No
such fast rotation period has been observed in any FRB; even FRB 200428,
known to be produced by a neutron star with a rotation period of 3.245 s,
has not shown that period in its FRB activity.

In each of these models detection of the rotation period would constrain
the other parameters; the cited original papers discuss these relationships.  
In this subsection I eliminate unknown parameters in favor of the observable
time derivative of the precession rate that may soon either be measured or
significantly constrained.}
\subsubsection{Free Precession}
\label{freepre}
Free precession of an aspherical neutron star deformed by magnetic stress
was suggested by \citet{LBB20,ZL20}.  In the model of \citet{LBB20} (the
model of \citet{ZL20} leads to similar conclusions) the precession angular
frequency $\Omega_{pre}$ varies
\begin{equation}
	\label{dOLBB}
	{\dot \Omega_{pre}} = - {\Omega_{pre} \over 2 t_{sd}},
\end{equation}
where $t_{sd}$ is the spindown age of the neutron star in FRB 20180916B
extrapolated back to infinite $\Omega_{pre}$\footnote{\citet{LBB20}
identify $t_{sd}$ with its actual age, but the neutron star could have been
born with spin and precession rates close to their present values.}.  Its
actual age must be less than $t_{sd}$.  The formulation of Eq.~\ref{dOLBB}
removes any explicit dependence on the magnetic parameters or rotation rate;
these are implicit in $t_{sd}$ and $\Omega_{pre}$.  The phase deviation from
a constant $\Omega_{pre}$ and period $P_{pre} = 2\pi/\Omega_{pre}$
\begin{equation}
	\label{Dphi}
	\Delta \phi = {1 \over 8} {\dot \Omega_{pre}} T^2 = - {\pi \over 8}
	{T^2 \over P_{pre} t_{sd}}.
\end{equation}

The data span 49 cycles ($T \approx 7 \times 10^7\,$s) and $|\Delta \phi|
\lesssim 0.3\,$radian may be estimated from \citet{Pl20,PM20}, leading to a
lower bound:
\begin{equation}
	\label{tmin}
	t_{sd} = {\pi \over 8}{T^2 \over P_{pre}\,
	\Delta\phi} \gtrsim 5 \times 10^9\,\text{s} \sim 150\,\text{y}.
\end{equation}
Because of its steep dependence on $T$, this lower bound on $t_{sd}$ may
increase rapidly as observations continue.  If, instead, a non-zero value of
$\Delta\phi$ were measured then Eq.~\ref{tmin} would provide a specific
value for $t_{sd}$.

However, no value of nor lower bound on $t_{sd}$ could exclude this model
because $t_{sd}$ is only an upper bound on the age of the object.  The
source's actual age is less, perhaps much less, than $t_{sd}$ if the neutron
star was born with spin and precession rates close to their present values.
The only possible upper bound on $t_{sd}$ would be $\sim T$, found if
$\Omega$ decreased by a factor ${\cal O}(1)$ over that time; this
would not provide any additional information because, by assumption, the
source would have been observed over the time $T$.

{The classic expression for the age of a rotating dipole, initially
rotating very fast, is
\begin{equation}
	\label{tsd}
	t_{sd} = {3 \over 4}{I c^3 \over \sin^2{\theta} \mu^2}
	\Omega_{rot}^{-2},
\end{equation}
where $I$ is its moment of inertia, $\theta$ the angle between rotation and
dipole axes, $\mu$ the magnetic moment and $R$ the radius.  In the model of
\citet{LBB20} this may be used to rewrite Eq.~\ref{tmin} in terms of the
polar dipole field $B = 2 \mu/R^3$:
\begin{equation}
	\label{BLBB20}
	B = \sqrt{{6 \over \pi^3}{I c^3 \epsilon^2 P_{pre}^3 \Delta \phi
	\over \sin^2{\theta} R^6 T^2}}.
\end{equation}
where the dynamical asymmetry $\epsilon = k (B_{int}/10^{18}\,\text{G})^2$
\citep{LBB20} and $k$ is a dimensionless parameter presumably ${\cal O}(1)$.
$B_{int}$ is a measure of the internal magnetic field that produces the
dynamical asymmetry, possibly $\gg B$ but not $\ll B$.

If $B_{int} \sim B$ then Eq.~\ref{BLBB20} may be rewritten, using the
definition of $\epsilon$,
\begin{equation}
	\label{BLBB20min}
	B \sim {10^{36}\,\text{G}^2 \over k} \sqrt{\sin^2{\theta} R^6
	\pi^3 T^2 \over 6 I c^3 P_{pre}^3 \Delta\phi} = 5 \times 10^{14}
	{\sin{\theta} \over k} \sqrt{0.3 \over \Delta\phi}
	{T \over \text{1 y}}\ \text{G},
\end{equation}
where $I = 10^{45}\,$g-cm$^2$ and $R = 10^6\,$cm have been assumed.
Because only an upper limit to $\Delta\phi$ is known, this last expression
is only a lower bound on $B$ that will increase with $T$ as data accumulate.
It is an indication of the values (if a nonzero $\Delta\phi \propto T^2$ is
measured) or bounds that may be placed by future observations.  It is now
consistent with expectations for strongly magnetized neutron stars, such as
the sources of Soft Gamma Repeaters (SGR), but may exceed such expectations
in the future.
}
\subsubsection{Radiatively Driven Precession}
\label{radpre}
In the radiatively driven precession model of \citet{So20} the rotation
period $P_{rot}$ and precession period $P_{pre}$ are proportional:
\begin{equation}
	{P_{rot} \over P_{pre}} = \left({B \over 7.45 \times 10^{17}\,
	\text{G}}\right)^2,
\end{equation}
where $B$ is the polar (not equatorial, the more usual parameter) field of a
dipole.  Combining this with the spindown equation
\begin{equation}
	{dP_{rot} \over dt} = {2 \over 3} (2\pi)^2
	{\mu^2 \sin^2{\theta} \over I c^3 P_{rot}},
\end{equation}
where the magnetic dipole moment $\mu = B R^3/2$, yields
\begin{equation}
	\label{dOS}
	{\dot \Omega_{pre}} = {2 \over 3} (2\pi)^3 {R^6 \sin^2{\theta} \over
	I c^3 P_{pre}} {(7.45 \times 10^{17}\,\text{G})^4 \over (2B)^2},
\end{equation}
where $R$ is the neutron star's radius, $I$ its moment of inertia and
$\theta$ the angle between the angular momentum and magnetic dipole axes.
This form replaces the explicit dependence on the unknown $P_{rot}$ with a
dependence on $B$ and the observed $P_{pre}$.  This can be used to relate
$B$ to a phase shift $\Delta\phi$ observed over a time $T$, using
Eq.~\ref{Omdot}:
\begin{equation}
	B = {(2\pi)^{3/2} \over 4\sqrt{3}}{T R^3 \sin{\theta} \over
	\sqrt{I c^3 P_{pre}^3}}{(7.45 \times 10^{17}\,\text{G})^2 \over
	\sqrt{\Delta\phi}}.
\end{equation}

For the conventional values $R = 10^6\,$cm and $I = 10^{45}\,$g-cm$^2$, the
polar dipole field
\begin{equation}
	B = 6.4 \times 10^{13} \sin{\theta} {T/\text{1 y} \over
	\sqrt{\Delta\phi}}\ \text{G} \gtrsim 2.5 \times 10^{14}
	\sin{\theta}\ \text{G},
\end{equation}
where the present $T = 2.2\,$y and $|\Delta\phi| \lesssim 0.3$ radian have
been taken.  Even for the maximum possible $\sin{\theta} = 1$ this bound is
consistent with the magnetic fields estimated for SGR, {and is
numerically similar to the corresponding Eq.~\ref{BLBB20min} for the free
precession model of \citet{LBB20,ZL20}}.  It cannot become large enough to
exclude the radiative precession model of \citet{So20} on the basis of
requiring an implausibly large field, even assuming $\sin{\theta} = 1$, for
many years (depending on how large a field is considered implausible).

{Just as in Sec.~\ref{freepre}, a value of $t_{sd}$ inferred from $B$
cannot by itself exclude or constrain this model.}
\subsubsection{Geodetic Precession}
\citet{YZ20} suggest periodic modulation of observed activity is produced
as the spin axis of a rotating neutron star precesses in the gravity of a
binary companion in a misaligned orbit.  This relativistic effect (geodetic
precession) does not depend on any asphericity of the neutron star.  Bursts
emitted in a cone around the spin axis may be directed towards the observer
only during some fraction of the precession cycle.  As the orbit decays by
emission of gravitational radiation the precession rate $\Omega_{pre}$
increases, producing a phase advance $\Delta\phi$.  The observational
constraint \citep{Pl20,PM20} on any such advance constrains the orbital
parameters.

Using the standard results for the shrinkage of the orbit, taken to be
circular with radius $a$,
\begin{equation}
	{da \over dt} = -{64 \over 5}{(GM_1)^3 \over c^5}
	{q(1+q) \over a^3},
\end{equation}
where $M_1$ is the mass of the source of the FRB and the mass ratio
$q = M_2/M_1$, and for the
precession rate
\begin{equation}
	\label{Oa}
	\Omega_{pre} = {(4 + 3q)q \over 1 + q}{GM_1 \over 2 a c^2}
	\Omega_{orb} = {(4 + 3q)q \over \sqrt{1 + q}}{(GM_1)^{3/2} \over
	2 a^{5/2} c^2},
\end{equation}
we find
\begin{equation}
	{\dot \Omega_{pre}} = 32 {(GM_1)^3 \over a^4 c^5} q(1 + q)
	\Omega_{pre}.
\end{equation}
Using Eq.~\ref{Oa} to find $a$ in terms of the observed period $P_{pre}$
and $\Omega_{pre} = 2\pi/P_{pre}$
\begin{equation}
	\label{Oadot}
	{\dot \Omega_{pre}} = 32 {2^{8/5}(1+q)^{9/5} \over q^{3/5}
	(4+3q)^{8/5}}{(GM_1)^{3/5} \over c^{9/5}} \Omega_{pre}^{13/5}.
\end{equation}
The explicit dependence on the unknown $a$ and $\Omega_{orb}$ has been
eliminated.  The dependence on $M_1$ and $q$ is not steep: the second factor
is 0.47 for $q=1$ and approaches $(2/3)^{8/5}q^{-2/5}$ as $q \to \infty$, as
shown in more detail by \citet{YZ20}, and in this model $\Omega_{pre}$ is
the observed modulation frequency.  

This model makes a testable quantitative prediction.  Taking $M_1 =
1.4M_\odot$ and $q = 1$
\begin{equation}
	{\dot \Omega}_{pre} = 1.5 \times 10^{-16}\ \text{s}^{-2}
\end{equation}
and
\begin{equation}
	\Delta \phi = {1 \over 8} {\dot \Omega} T^2 = 0.018 \left({T \over
	\text{1 y}}\right)^2\ \text{radian}.
\end{equation}
With the present $T \approx 2.2\,$y the predicted phase drift is approaching
the accuracy of measurement.  The prediction will soon be tested (provided
the FRB remains active).  Note that $\Delta \phi$ is four times larger if
the period is accurately determined at one end of the data duration $T$
rather than fitted to the entire duration with uniform weight; this may be
applicable if detected bursts are non-uniformly distributed in time.
\subsubsection{Newtonian Precession by a Fall-back Disc}
\citet{TWW20} suggest the direction of FRB radiation is modulated by the
precession of a neutron star under the Newtonian gravitational influence of
a fall-back disc.  This model makes the qualitative prediction that the
modulation period should increase as the disc dissipates, some matter
accreting onto the neutron star, possibly making SGR and Anomalous X-Ray
Pulsar (AXP) emission \citep{KTU94}, and some driven to greater distances.
If this occurs on a characteristic time scale $t_{diss}$ then
${\dot \Omega}_{pre} \sim -\Omega_{pre}/t_{diss}$.

The dissipation time is unknown; guessing $t_{diss} \sim 10^{11}\,$s from
the ages of SGR (inferred from the ages of the supernova remnants
containing them), suggests ${\dot \Omega}_{pre} \sim 4 \times 10^{-17}
\,$s$^{-2}$.  A phase shift $\Delta\phi = 1\,$radian, unambiguously
detectable, would occur in a time $T \sim \sqrt{8/{\dot \Omega_{pre}}} \sim
15\,$y.

Any quantitative prediction, as well as the model itself (almost nothing is
known about fall-back discs beyond the evidence that many white dwarfs and
at least a few neutron stars are surrounded by some orbiting material)
depends on several speculative assumptions.  Even the sign of ${\dot
\Omega}_{pre}$
is uncertain, because matter may move inward, increasing $\Omega_{pre}$, as
well as moving outward or being lost, decreasing $\Omega_{pre}$.
\subsection{Neutron Star Rotation}
\citet{BWM20} suggested that the 16 day period may be the rotation period of
a neutron star.  The longest known neutron star rotation periods in X-ray
binaries are about 1000 s, about a thousand times shorter that that of FRB
20180916B.  1E 161348$-$5055 has been suggested \citep{DL06} to be a single
neutron star rotating with a 6.67 hour period, about 50 times shorter than
that of FRB 20180916B.

A neutron star, estimated to have $I \approx 10^{45}\,$g-cm$^2$, accreting
freely infalling matter at its surface, would spin up from rest to a 16 day
period by the accretion of $10^{-10}$ of its mass, provided the angular
momentum of the accreted mass were all aligned.  Accretion from a disc would
require only $1/\sqrt{2}$ as much mass.  Interaction of either accretion
flow with a magnetosphere at radius $10^9\,$cm would reduce these mass
fractions by a factor of about 30.  Neutron stars interacting with the
winds of binary companions appear to be spun down by a ``propellor effect''
(usually when they are not accreting; they usually spin up when accreting)
to periods as long as hundreds of seconds.

{These considerations can be restated in terms of bounds on any
systematic torque $N$, of accretional or other origin, placed by the
observed bound Eq.~\ref{Omdot} on the rate of change of the modulation
frequency:
\begin{equation}
	N \lesssim {8 \Delta\phi \over T^2} I \sim 5 \times 10^{29}
	\text{dyne-cm}.
\end{equation}
If interpreted as the result of accretion onto the surface of a neutron
star, the corresponding upper bound on the accretion rate $\sim 3 \times
10^{13} \text{g/s} \approx 5 \times 10^{-13}M_\odot/\text{y}$.  This is much
less than accretion rates in known massive X-ray (NS-OB star) binaries.}

Although not demonstrably impossible, a single neutron star with a spin
period of 16 days may be implausible.  The existence of single fast young
pulsars like the Crab suggests rapid (periods of tens of ms) initial spin.
If collapsed from a stellar core of white dwarf density, conserving angular
momentum, a neutron star with a 16 day rotation period would imply a
pre-collapse core's rotation period of thousands of years.  Even the 6.67 h
period of 1E 161348$-$5055 might have other explanations, such as the
orbital period if it has a low mass companion.

The AM Her stars (``polars'') may be informative.  The rotation of these
binary magnetic white dwarves is synchronous with their orbits.
Synchronism is produced by magnetostatic interaction \citep{JKR79}, as
demonstrated by the preferential orientation \citep{C88} of the white
dwarf's magnetic dipole in the rotating frame \citep{K89}.  A similar
phenomenon might make FRB 20180916B rotate synchronously with a
companion star in a 16.3 d orbit, but the maximum synchronizing torque would
be small:
\begin{equation}
	\label{Nmax}
	N_{max} \sim {\mu^2 \over a^3},
\end{equation}
where $\mu \approx 10^{33} B_{15}\,$G-cm$^3$ is the neutron star's magnetic
moment.  For a 16 day period $a \sim 10^{12}\,$cm and $N_{max} \sim 10^{30}
B_{15}^2\,$dyne-cm.

Synchronism would be disrupted by an accretion rate $\gtrsim 10^{-13}
M_\odot$/y, but might be maintained if the accretion rate were very small.
Eq.~\ref{Nmax} also sets an upper limit to the synchronizing torque,
achieved only if the effective coupling is strongly dissipative ($Q \sim 1$
if considered as an oscillator).  Synchronization might be possible in a
time $t$ if the initial neutron star rotation rate $\Omega_0 < N_{max}t/(IQ)
\sim [t/(3 \times 10^4 Q \text{y})] B_{15}^2\,\text{s}^{-1}$.  The implied
magnetic field at the companion would be $\sim$mG, roughly comparable to
that inferred over a larger distance scale \citep{K21} for the environment
of FRB 121102.  The maximal accretion rate consistent with synchronism also
corresponds to an electron density comparable to that estimated for the
magnetoionic environment of FRB 121102.  However, FRB 20180916B does not
show the large and rapidly varying rotation measure of FRB 121102, so it is
unclear if the parameters inferred for one of these objects can apply to the
other.
\section{Accretion Disc Precession}
\label{disc}
Many X-ray binaries \citep{SFOND,TC20} and cataclysmic variables \citep{A13}
show superorbital periods explained as precession of the axes of accretion
discs misaligned with the orbital plane \citep{K73,K80}.  These periods are
tens to a few hundred days, consistent with the period of FRB 20180916B.
This model was confirmed by observation of sideband periods in SS 433
\citep{KAMG} and of the phases of the disc eclipses in Her X-1 \citep{LJ82}.
Alternative models have been discussed by \citet{ZPS11} and are not favored.

Misalignment of the disc and orbital planes can be considered a perturbation
to the lowest energy state of a disc, in which it lies in the orbital plane.
Precession is observed in many, if not all, of those binaries that are
observed at large inclination so that the disc may eclipse the compact
object or the central, high-energy emitting, regions of the disc itself (in
SS 433 measurement of the Doppler shifts of the jets makes eclipse
unnecessary).

The mechanism of excitation is unclear.  Speculations have included
radiative feedback from the outer disc shadowing the mass-losing companion.
A companion with misaligned spin would directly feed a misaligned disc, but
spin alignment is expected to occur at least as fast as orbital
circularization, and where accurately measured (as in Her X-1) the orbits
are accurately circular.  If the mass-losing star does have a misaligned
spin, then the rate of precession of the disc would be the sum of its
precession rate under the companion's torque and the precession rate of the
companion under the torque of the compact object.  The second term will
generally be smaller because the mass of most stars, especially in their
mass-losing phase of advanced evolution, is centrally concentrated.

In this model of the 16.3 d period of FRB 20180926B based on precession of
an accretion disc in a mass-transfer binary the bursts must be confined
within a cone or funnel around the disc axis \citep{K17} whose width
$\delta\theta$ is less than the amplitude of precession $\Delta\theta$.
Then $2\delta\theta/(2\pi\Delta\theta) \approx F$, where $F \approx 0.3$ is
the fraction of the 16.3 d cycle over which bursts are observed.  This does
not provide any clue to the values of $\delta\theta$ and $\Delta\theta$,
other than their ratio.

The existence of FRB 200428, a repeating FRB for which there is no evidence
of periodic modulation, produced by a single non-accreting neutron star,
is not evidence against models of modulated FRB that involve binary
companions or accretion discs if coherent emission occurs in a wide range of
environments.  The existence of short and long gamma-ray bursts, with
similar phenomenology but produced by different astronomical events, is
another example of phenomenological universality.
\section{Discussion}
Models may be distinguished by the behavior of the modulation period.
Orbital periods and orbitally-locked rotation periods involve the motion of
massive bodies and are expected to be extremely stable.  An exception to
this is geodetic spin precession in very compact binaries, whose period
shortens at a quantitatively predictable rate (Eq.~\ref{Oadot}).  This rate
is only weakly dependent on unknown parameters, and the model is therefore
testable.  Free and radiative precession of an isolated neutron star is
expected to slow smoothly on the star's spin-down time scale, although their
rates of decline depend on poorly known parameters (Eqs.~\ref{dOLBB},
\ref{dOS}).

Models involving fall back or accretion discs make no specific predictions
and are sensitive to unknown parameters (Eq.~\ref{dOS}).  These models,
unlike others, are consistent with irregular variation of the period.  In
any model involving disc dynamics the period depends on the mass
distribution in the disc, whose behavior is not known, and may fluctuate.
Observations of the best-studied accretion discs, those around Her X-1 and
in SS 433, indicate their periods vary slightly, with $\Delta P/P \sim
0.01$, without a systematic trend.  This is in contrast to precession of a
neutron star driven by a fall-back disc, whose period would be expected to
increase as the disc dissipates, but on the (unknown) time scale of the age
of the neutron star and disc.

The measured nonzero changes in RM and upper bounds on changes in DM
\citep{Pl20} of FRB 20180916B imply that the magnetic field in the region in
which the RM change is produced $B \gtrsim 3\,\mu$G.  This is consistent
with interstellar fields, and, unlike FRB 121102 \citep{K21}, does not
require an extraordinary local magnetoionic environment.
\section*{Data Availability}
This theoretical study did not obtain any new data.

\label{lastpage} %MNRAS
\end{document}